\documentclass[runningheads]{llncs}

\usepackage[T1]{fontenc}
\usepackage[export]{adjustbox}
\usepackage{graphicx}

\usepackage{tikz}
\usetikzlibrary{automata,positioning,fit,calc,backgrounds,shapes,arrows}
\usepackage{array,multirow,graphicx}
\usepackage{amssymb} 

\usepackage{hyperref}

\usepackage[citestyle=lncs,backend=biber,doi=true,url=true,isbn=false,giveninits=true,defernumbers=true,maxnames=20,maxcitenames=20]{biblatex}
\newcommand{\logikey}{\textsc{LogiKEy}}

\usepackage{xcolor}

\begin{document}

\title{Many Logics, One Methodology: A Plea for Logical Pluralism in Formalised Reasoning (preprint)}

\titlerunning{Many Logics, One Methodology}

\author{Christoph Benzmüller\inst{1,2}\orcidID{0000-0002-3392-3093} 
\and Daniel Kirchner\inst{1}\orcidID{0000-0001-9229-1148} 
\and Luca Pasetto\inst{3}\orcidID{0000-0003-1036-1718}
}

\institute{Otto-Friedrich-Universit\"at Bamberg, Kapuzinerstraße 16, 96047 Bamberg, Germany \and
Freie Universit\"at Berlin, Kaiserswerther Str. 16-18,
14195 Berlin, Germany \and
University of Luxembourg, P. de l’Université 2, 4365 Esch-sur-Alzette, Luxembourg\\
}

\maketitle              

\begin{abstract} 

This position statement looks back on two decades of work on shallow embeddings of non-classical logics in classical higher-order logic (HOL), a line of research that expanded into a range of logic embeddings in HOL and inspired the {\logikey} logic-pluralistic knowledge representation and reasoning methodology. This paper advances the case for logical pluralism at object-logic level within a unifying meta-logical framework such as \logikey, grounding the argument in computational metaphysics. More broadly, it advocates principled support for logical pluralism in modern proof assistants, and cautions against logical imperialism—the rigid adoption of a single foundational logic for large-scale theory developments—which impedes the interdisciplinary reuse that {\logikey} is designed to enable.

\end{abstract}

\section{Introduction}
The work reported here belongs to a research line, now spanning roughly two decades, on the development, formalisation, and automation of shallow embeddings of non-classical logics in classical higher-order logic (HOL). This line grew out of a side initiative within the Leo-II higher-order theorem-prover project \cite{J30,C26}\footnote{Leo-II, the winner of the CASC competition in 2010 in the higher-order (THF0) category, was developed with funding from EPSRC grant EP/D070511/1.} around 2007---at a time when the automated-reasoning community was still primarily focused on less expressive first-order and propositional logics, and had yet to appreciate the long-term importance of building systems capable of handling expressive higher-order classical and non-classical logics, which are inevitably needed to enable computer-supported studies on foundational topics in metaphysics and (meta-)mathematics.

The line began with a first joint paper (of the first author, with Paulson) on the embedding of multi-modal logic in HOL \cite{B9}, subsequently extended and superseded by two journal articles: \cite{J21}, on propositional multi-modal and intuitionistic logics in HOL, and~\cite{J23,R45}, on higher-order quantified multi-modal logics in HOL.

Building on these articles, a wide range of logic embeddings in HOL were developed, including access control logic, quantified conditional logics, multivalued logic, free logics, various deontic logics, intensional HO modal logics, and public announcement logic; see also \cite{J41,J48} and the references therein. These articles subsequently inspired what was introduced and developed in \cite{J48} as the {\logikey} logic-pluralistic knowledge representation and reasoning methodology; cf.~Fig.~\ref{fig:LogiKEy} for an instantiation of the {\logikey} methodology for the application direction addressed in this paper.
Publication \cite{J23} was particularly instrumental in enabling fruitful applications in computational metaphysics, including first studies~\cite{C40,C55} on Gödel's \cite{GoedelNotes} and Scott's \cite{ScottNotes} modal variants of the ontological argument. 
A recent article \cite{J75}, co-authored by Benzmüller and Scott 
in a special issue of Monatshefte für Mathematik dedicated to Kurt Gödel, provides a comprehensive discussion of these applications alongside some novel results and pointers to earlier and related works on this topic.

This paper---a position statement, enriched by some novel results---revisits this line of work, reflects on its development, and points toward promising directions for future research. It makes a case for logical pluralism at the object-logic level, pursued within a unifying meta-logical framework.

To exemplarily ground the proposal in a concrete application, the paper argues that the possible-world semantics as, for example, employed in the formalisation of Gödel's modal ontological argument can and should be naturally extended with additional logical and mathematical notions, allowing modalised mathematical theories and Gödel's modal ontological theory to coexist within the sketched unified logical framework; cf.~Fig.~\ref{fig:LogiKEy}. This unification—and the study of its implications for the modal ontological argument—is rendered particularly compelling by Gödel's well-documented mathematical realism, according to which mathematical objects enjoy an existence wholly independent of human cognition. Once the existence of the generally infinite objects and structures of mathematics is granted, Gödel's modal ontological theory is confronted from the outset with infinitely many positive properties, with the consequence that trivial finite models—of the kind presented in various contributions to the literature on the modal ontological argument—are thereby excluded. The notion of positive properties in Gödel's theory will in fact be forced to be uncountably infinite (i.e.~there must exist uncountably many distinct positive properties), an observation that deserves further investigation and for which the present paper offers a starting point.

From a broader perspective, this paper aims to illustrate the need for principled support for \emph{logical pluralism} (cf.~\cite{sep-logical-pluralism} and the references therein) within modern proof assistant systems—as opposed to what one might call \emph{logical imperialism}, where a particular foundational logic is adopted as the unquestioned and immovable basis for large-scale theory developments. Such rigidity risks hindering the reuse of existing libraries for interdisciplinary research of the kind sketched here, where mathematical theories must interface with other disciplines that do not share—or actively reject—the foundational assumptions underlying those libraries.

The remainder of this paper is organised as follows. Section~\ref{sec:logikey} introduces the \logikey\ methodology in more detail, contrasts logical pluralism with what we call logical imperialism, and positions \logikey\ relative to Zalta's Principia Logico-Metaphysica (PLM) and to Isabelle's own original pluralistic design. Section~\ref{sec:godel} then illustrates the methodology with a concrete application: starting from a shallow embedding of higher-order modal logic (HOML) in classical higher-order logic (HOL), modalised mathematical notions are introduced and combined with Gödel's modal ontological argument, leading to (first-time formalised) cardinality results about the set of positive properties — culminating in its uncountability. Section~\ref{sec:conclusion} concludes the paper. The Isabelle/HOL source files associated with this paper are contained in the \LaTeX-sources of the arXiv preprint package.\footnote{Self-contained excerpts of the Isabelle/HOL formalisations underlying the discussion are collected in Figures~\ref{fig:HOMLinHOL}--\ref{fig:GoedelVariantHOML2}; the reader is encouraged to consult Fig.~\ref{fig:HOMLinHOL} (the embedding of HOML in HOL) early on, as the notation introduced there—in particular the syntactic distinction between possibilist (\(\mathbf{\forall},\mathbf{\exists}\)) and actualist (\(\mathbf{\forall}^E,\mathbf{\exists}^E\)) quantifiers, the lifted modal connectives, Leibniz equality \(\equiv\), and global validity \texttt{Mvalid}—is used throughout Section~\ref{sec:godel}.}

\section{{\logikey}: Logical Pluralism vs. Logical Imperialism}
\label{sec:logikey}

\tikzset{block/.style={draw, thick, text width=2cm, minimum height=1.3cm, align=center}, 
         line/.style={-latex} }
\tikzset{ font={\fontsize{12pt}{12}\selectfont}}

\tikzset{testpic1/.pic={ 
\node[block, fill=green!20, text width=21cm, minimum width=21cm, minimum height=3cm] (m1) 
      { \textbf{\huge L3 --- Applications (using L2):  \\[.5em] Further Studies on Gödel's Modal Ontological Argument} };
\node[block, below=.5cm of m1, fill=orange!30, text width=21cm, minimum height=3cm] (m2) 
      { \textbf{\huge L2 --- Domain-specific Theories (modeled in L1): \\[.5em] Gödel's Modal Ontological Theory enriched with Modal Maths} };
\node[block, below=.5cm of m2, fill=yellow!20, text width=17cm, minimum width=21cm, minimum height=3cm] (m3) 
      { \textbf{\huge L1 --- object-logic(s) (embedded in L0): \\[.5em] Higher-Order Modal Logic (HOML)} };
\node[block, below=.5cm of m3, fill=blue!20, text width=17cm, minimum width=21cm, minimum height=3cm] (m4) 
      { \textbf{\huge L0 --- Meta-Logic: \\[.5em] Classical Higher-Order Logic (HOL)} };
\node[block, above right=-.5cm and 1.5cm of m1, fill=gray!40, text width=12.5cm, minimum width=9cm, minimum height=2cm, rotate=-90] (r1) 
      { \textbf{\huge \logikey\ Methodology} };

  \begin{scope}[on background layer] 
    \node[draw, fill=blue!20, inner xsep=5mm, inner ysep=5mm, fill opacity=0.6, fit=(m1)(m2)(m3)(m4)](m1tom4){};
  \end{scope}

  \begin{scope}[on background layer] 
    \node[draw, fill=gray!40, inner xsep=20mm, inner ysep=5mm, fill opacity=0.5, fit=(m1)(m2)(m3)(m4)(m1tom4)](all){};
  \end{scope}
}}

\begin{figure*}[tp!]
\centering 
\begin{minipage}{.80\textwidth}
\resizebox{\textwidth}{!}{
    \begin{tikzpicture}
      \pic{testpic1};
    \end{tikzpicture}
    }
\end{minipage}
\caption{The logic-pluralistic knowledge representation and reasoning methodology \logikey\ instantiated for the application direction proposed in this paper. \label{fig:LogiKEy}}
\end{figure*}
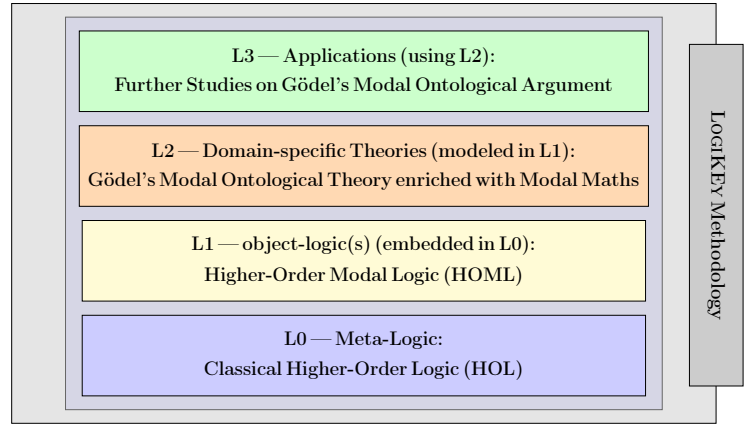

\subsection{The \logikey\ methodology\label{sec:logikey-methodology}}

The term \logikey, introduced in \cite{J48}, refers to a logic-pluralistic knowledge representation and reasoning methodology and infrastructure that has in recent years been frequently deployed with Isabelle/HOL \cite{Isabelle} as its host environment—benefiting from its simultaneous support for interactive proof, proof automation, and (counter-)model finding. \logikey\ is the acronym of ``\textbf{Logi}c and \textbf{K}nowledge \textbf{E}ngineering Framework and Methodolog\textbf{y}''. The methodology applies logic-based knowledge representation and reasoning to engineering tasks in which the underlying object-logic itself is a negotiable part of the design space. More concisely, \emph{the \logikey\ methodology proceeds by first embedding an object-logic of interest---e.g.,~a higher-order modal, deontic, conditional, or free logic---inside a meta-logic (here classical HOL) by interpreting its formulas as world- or, more generally, context-relativised predicates in HOL; domain-specific theories are then formalised on top, and applications on top of those, yielding a layered architecture (cf.~Fig.~\ref{fig:LogiKEy}; the layers include, but are not limited to: L0---meta-logic, L1---object-logic(s), L2---domain theories, and L3---applications)}.\footnote{This historically meant \emph{shallow} embeddings; the methodology, however, is not committed to that choice (see the discussion at the end of this subsection).} Because the embedding lives inside HOL, the host system's full automation, model-finding, and proof-reconstruction infrastructure is available at every layer, and the object-logic itself becomes a negotiable parameter that can be explicitly analysed, revised, exchanged, compared, combined, etc.---exactly the pluralistic stance this paper advocates.
The \logikey\ methodology is not bound to Isabelle/HOL, however, and any proof assistant based on a sufficiently strong logic can serve as host instead. The initial embedding work and the early studies on Gödel's modal ontological argument \cite{C40,C55} used the ATP Leo-II \cite{J30}, which in turn influenced the built-in support for logical pluralism in its successor Leo-III \cite{J72,J51}. Early studies using Rocq (formerly Coq) \cite{Coq} as host have also been carried out~\cite{C44}.
Beyond computational metaphysics, the \logikey\ methodology has more recently been applied to logics and formalisms for normative and legal reasoning \cite{J48}. A key feature across both areas is that within the HOL meta-logic, object-logics—such as higher-order (multi-)modal logics or deontic logics—become first-class objects of study: they are developed, examined, and refined in direct interaction with the application domain.

\paragraph{Shallow embeddings, adequacy, and consistency.}

Two concerns about the approach deserve acknowledgement.
The first is that the embeddings used within \logikey\ have historically been
\emph{shallow}, in the sense that object-logic formulas are identified with
HOL terms (e.g., via the standard translation). Such an embedding is,
strictly speaking, a model construction rather than a syntactic representation
of the object-logic, so one loses access to logic-specific
syntactic tooling. 
However, (i) in exchange one gains the full automation stack of a mature classical higher-order prover—Sledgehammer, Nitpick, Metis, SMT bridges, and the like—which empirically scales to substantial formalisations; and (ii) the methodology is in fact not committed to the shallow choice: recent work~\cite{Benzmueller2025FaithfulDeepShallow,BenzmuellerKirchner2026FMLDeepShallow} shows that a \emph{deep} embedding (as an inductive datatype of formulas) can be developed alongside the shallow one inside the same Isabelle/HOL theory, with mutual faithfulness proofs mechanised and largely automated (demonstrated so far for propositional and first-order modal logic). The same
layered \logikey\ architecture accommodates shallow and deep variants---and
indeed their combination---at the L1 level. Lifting this
\emph{deep-and-shallow} methodology to HOML, and onward to further (existing
and novel) object-logics in the \logikey\ portfolio, is a high priority
for future work. The second concern is that combining HOL with
object-logic-specific axioms yields a hybrid system whose relative consistency
does not follow directly from either component---a familiar issue from
HOLZF~\cite{Obua2006HOLZF} and Paulson's \emph{ZFC in
HOL}~\cite{Paulson_ZFC_in_HOL_AFP}. In our setting, however, HOML is
introduced \emph{purely by definitions}, with any axiomatic content
(e.g.,~Gödel's axioms in Fig.~\ref{fig:GoedelVariantHOML2}) layered on top and
open to inspection (cf.~Sect.~\ref{sec:godel}); relative consistency therefore
reduces in the standard way. Soundness and completeness of the shallow
embeddings have, moreover, been established by pen-and-paper proofs for many \logikey\
logics (cf.~\cite{J21,J23} and the broader \logikey\ literature); what
has so far been missing is a \emph{mechanisation} of such results
inside Isabelle/HOL itself---which the extended deep-and-shallow \logikey\
methodology now initiates.

\subsection{Logical imperialism: monoculture and invisible assumptions\label{sec:imperialism}}

The \logikey\ methodology stands in contrast to the prevailing tendency in the development of mathematics libraries in modern proof assistants—a tendency we have somewhat provocatively termed logical imperialism above. Such systems implicitly privilege a particular logical or foundational framework—based, for instance, on dependent type theory, constructive logic, first-order set theory, or classical HOL—as the default or sole correct basis for all formal reasoning. The consequences are threefold: it fosters \emph{monoculture by design}, where adopting an alternative foundation incurs significant encoding overhead; \emph{cultural dominance}, where it becomes increasingly difficult to even ask ``what if we used a different logic?''; and it may embed \emph{invisible assumptions}—such as the law of excluded middle, existential import, the axiom of choice, impredicativity, extensionality, the treatment of undefinedness, and others—directly into the system, rendering them largely opaque to the user.

One might argue that the recent successes of generative AI offer an easy remedy, since future generations of such systems could eventually assist in adapting existing large libraries to alternative logical foundations. While this may well prove true, it does not count against the \logikey\ methodology—on the contrary, when used in combination with this approach, such adaptations should be even better and more transparently supported, given the availability of explicit object-level--meta-level connections as formal data.

\sloppy Isabelle was originally designed by Paulson with logical pluralism in mind~\cite{cs-LO-9301106}. It is a generic proof assistant framework built around a minimal meta-logic (Isabelle/Pure), on top of which different object-logics can be instantiated—Isabelle/HOL, Isabelle/ZF, and Isabelle/FOL being prominent examples.
It is worth noting that Isabelle's choice of a minimal meta-logic, over which object-logics are axiomatised, constitutes a key architectural difference from \logikey's preferred approach, which takes classical HOL as its meta-logic and treats different object-logics as naturally embedded substructures rather than axiomatically introduced ones. This distinction carries practical benefits—for proof automation in particular.

In the Isabelle landscape, Isabelle/HOL has come to dominate the ecosystem in practice, with the vast majority of libraries, tools, and community efforts built exclusively around it. In this sense, the original pluralistic vision of Isabelle has been partially set aside in practical applications—including, notably, the development of formalised mathematics libraries—even if it remains present in the underlying architecture. The situation is no better in other current systems and library developments, and is arguably worse in those that were never designed with logical pluralism in mind.

It must be acknowledged that, for pragmatic reasons, such foundational commitments must eventually be made in order to achieve meaningful progress on ambitious library projects. A prominent recent example is Mathlib \cite{mathlib2020}, built on Lean's \cite{Lean2021} dependent type theory—specifically, a variant of the Calculus of Constructions with universes—with global commitments to classical logic, the axiom of choice, and Boolean extensionality.
While this pragmatic stance enables the remarkable scale and coherence of such projects, it risks coming, as we fear, at the cost of foundational inflexibility and, in places, mathematically or philosophically questionable consequences.

\subsection{The hidden costs of pragmatic conventions\label{sec:hidden-costs}}

A telling example is the treatment of division by zero in combination with existential import. One pragmatically motivated choice is to stipulate  $1/0 = 0$ by definition, which makes statements such as  $\exists x.\ 1/0 = x$ and $\exists x. \forall  y.\ y/0 = x$ theorems. The first is merely odd; the second—asserting that all division-by-zero expressions collapse to a single, even ``existing'' object—would strike most philosophers of mathematics as foundationally untenable. From a Platonist or structuralist perspective, these are not mathematical truths but formal artefacts of a particular definitional choice. From a formalist or fictionalist perspective they are valid, but only within the chosen system—and that qualification is precisely what tends to remain invisible to future generations of library users.\footnote{It must be acknowledged, however, that this convention is
not without genuine pragmatic merit: a totally defined division operator
allows unconditional rewriting steps such as $(x+y)/z = x/z + y/z$ without the
side condition $z\neq 0$, and may thus streamline proof automation in large
mathematics libraries. Moreover, such conventions are easily translated
between proof assistants---HOL Light's $x/0 = 0$ and HOL4's ``undefined
arbitrary value'' relate via a suitable \textsf{if-then-else} wrapper---and
can be refined locally on top of a classical HOL library by a conditional
definition such as $y \neq 0 \Longrightarrow \mathit{div}\, x\, y = x/y$,
without modifying the underlying logic. Our concern is therefore not that such
conventions are illegitimate, but that, once built into libraries reused as if
they were neutral mathematical bedrock, they become invisible logical
assumptions; we plead for their visibility and revisability, not their
abolition. The case for logical pluralism is accordingly stronger in
foundational and philosophical contexts---such as the metaphysical
applications discussed below---than in many industrial verification settings,
where a single, well-understood, classically total foundation may be the most
effective engineering choice.}

Crucially, this need not be an inevitable trade-off. Logics such as free logic \cite{Lambert60,Scott67}—which can in fact be embedded in HOL in a straightforward manner~\cite{J40}—are specifically designed to handle partial functions and undefinedness in a principled way, without resorting to junk values. We stress that adopting such a logic is by no means the only way to keep
one's foundational options open (e.g., a conditional definition layered on
top of a total HOL operator may afford comparable flexibility within HOL
itself). The deeper concern is neither the particular convention adopted
nor the fact that pragmatic choices are made---such choices are arguably
unavoidable at the scale of a modern mathematics library---but rather
their explicit visibility and how they are received downstream. When future users—and future AI systems!—build on such libraries, there is a real risk that they will inherit their foundational commitments unreflectedly, treating them as neutral mathematical bedrock rather than as one considered option among several, each carrying its own philosophical commitments and formal consequences.

More pressing still is that foundational flexibility and logical pluralism are not merely desirable but necessary for certain interdisciplinary applications of formalised reasoning that seek to bridge distinct domains—say, mathematics with philosophy and metaphysics. In metaphysical studies such as those concerned with Gödel's modal ontological argument, principles routinely taken for granted in classical mathematics libraries must frequently be rejected outright to avoid trivialisation, paradox, and inconsistency. At the same time, the metaphysical structures under investigation often stand in deep and important relationships to cognate mathematical structures. A case in point is Gödel's notion of positive properties in his modal ontological theory, which corresponds to a modalised ultrafilter on sets, or equivalently on properties \cite{J52,J75}—an insight that calls for a unified framework in which both the metaphysical and mathematical dimensions can be investigated together.\footnote{Readers unfamiliar with these notions may wish to know in advance that ``positive properties'' is a primitive notion in Gödel's modal ontological argument, governed by axioms that, taken together, force this set of properties to behave as an ultrafilter in the modal setting; the precise formal counterpart---a \emph{modal} ultrafilter on the world-relativised power set---is defined in Section~\ref{sec:godel} and Fig.~\ref{fig:ModalFilter}, and the connection to Gödel's axioms is reviewed there. ``Positive'' here is the term Gödel inherits from Leibniz, and---in
Gödel's own words---is intended ``in the moral aesthetic sense (independently of the accidental structure of the world)''~\cite{GoedelNotes}; it has no connection to any of the technical uses of ``positive'' in logic or computer science.} This demand is further reinforced by Gödel's nuanced mathematical realism, which motivates enriching his modal ontological theory with foundational mathematical concepts and structures—e.g.~a theory of natural numbers—and which, as we argue, would thereby rule out trivial finite interpretations of positive properties. This provides yet another compelling reason to pursue a unified foundational framework that is both logically flexible and interdisciplinarily adequate.

\subsection{Principled monism: Zalta's PLM as alternative\label{sec:plm}}

At this point it is worth noting that philosophers have developed principled approaches aimed at proper foundations for all of the sciences, including pluralistic entry points for mathematics. Zalta's Principia Logico-Metaphysica (PLM) \cite{PLM} is one such serious and sophisticated system. It formalises abstract object theory (AOT) \cite{Zalta1983,Zalta1988} within a higher-order hyperintensional modal logic, with a carefully crafted treatment of encoding versus exemplification, existence, and undefinedness, based on a relational base logic, and it has been used to formalise a remarkable range of metaphysical results. Briefly, AOT distinguishes two modes of predication: ordinary objects \emph{exemplify} properties (the familiar predication of standard logic), while \emph{abstract} objects, in addition, may \emph{encode} properties---i.e.~have them as constitutive parts of their nature without instantiating them in the usual sense; this distinction underwrites Zalta's reconstructions of Fregean numbers, Platonic forms, situations, possible worlds, fictional and mythical objects, and Leibnizian concepts, among others (see~\cite{PLM} for the full catalogue). The system is also \emph{hyperintensional}, in the sense that necessarily equivalent properties are not automatically identified, which is essential for the metaphysical distinctions just listed.

A sophisticated challenge to the logical pluralism advocated here might therefore be posed as follows. Rather than keeping the object-logic negotiable, why not choose, in the spirit of logical monism, a single foundation rich enough that negotiation becomes unnecessary? PLM thus by no means represents crude or unreflective logical imperialism, but a principled response to it: a foundation specifically engineered to be adequate for both mathematics and metaphysics simultaneously.

We resist this conclusion, for two reasons. First, PLM's very richness is a liability as much as an asset: its foundational commitments are substantial and distinctive---the encoding/exemplification distinction, the choice of a relational base logic, the specific treatment of definite descriptions and undefinedness, and the strict hyperintensionality of property identity, to name only the most visible ones---and researchers who do not endorse this particular package (for instance, because they wish to remain neutral on hyperintensionality, prefer a functional-style base logic, or are exploring weaker or alternative metaphysical commitments) find themselves working against the grain of the system rather than with it. Second, and more fundamentally, PLM is a \emph{foundational theory}—a specific set of ontological and logical commitments advanced as the correct foundation—whereas \logikey\ is a \emph{methodology}, one that treats the choice of object-logic, and to some extent even the choice of meta-logic, as itself an object of study rather than a settled question.

A vivid illustration of this distinction comes from the second author's earlier work~\cite{KirchnerPhD}, in which the \logikey\ methodology---with appropriate adaptations---was used to embed the logical foundations of PLM and AOT as an object-logic within classical higher-order logic. This made PLM itself an object of formal study, with a notable outcome: a previously known paradox, which had been inadvertently reintroduced without detection, was identified through interaction with Isabelle/HOL and subsequently corrected. Far from undermining PLM, this episode exemplifies exactly what logical pluralism enables: the ability to step outside any given system, examine its foundations critically, and compare it with alternatives.

Furthermore, \logikey\ is not committed to a fixed meta-logical layer. The current choice of classical HOL rests on pragmatic grounds: it offers a concise and well-understood syntax and semantics, together with comparatively mature automation support relative to other logics of comparable expressiveness.\footnote{In fact, for reasons of purity and to support reuse, applications of \logikey\ so far have always aimed to stay as close as possible to Church's simple type theory as meta-logic, avoiding the use of additional built-in theories and mechanisms in Isabelle/HOL at the meta-logical layer wherever possible.} But this choice is not constitutive of the \logikey\ methodology. Conceptually, additional layers can be introduced, so that within an ultimate meta-logic—HOL, or something beyond it—expressive foundational logics and theories, including AOT and PLM itself, can be embedded as object-logics, which may in turn serve as meta-logics for yet further encodings. This regress is not vicious but productive: it is precisely what distinguishes a methodology from a foundation. A foundation forecloses; a methodology explores. PLM, in this sense, narrows the logical landscape by design—whereas \logikey\ maps it. That said, the dialogue between the two approaches has proven mutually illuminating, and there is no reason to regard them as adversaries rather than complements.

It is worth noting that the unifying formalisation tasks motivated and initiated in this work could in principle be carried out using PLM as a foundation---and related work connecting Gödel's theory with mathematical structures has indeed been presented recently by Zalta \cite{ZaltaUnifying2025}. The distinction discussed above nonetheless persists: the only currently available means of verifying such work on a computer is through the embedding of PLM in HOL described in~\cite{KirchnerPhD}, following the \logikey\ methodology. A dedicated, native proof assistant for AOT and PLM could, of course, be developed, but doing so would require considerable effort if undertaken manually---though, as noted above, generative AI may eventually prove helpful in this regard. Furthermore, the broad foundational ambitions of AOT and, in particular, its fine-grained hyperintensionality and its careful treatment of philosophical nuances in modal reasoning\footnote{E.g. AOT distinguishes between \emph{modally-strict} reasoning and reasoning from necessities that may be consequences of contingent axioms.} makes meaningful automation support challenging to achieve, at least in comparison to HOL. The point here is methodological rather than principled: AOT is presented as a list of axioms over a relational, hyperintensional base logic, and crucial reasoning steps---e.g.~moving between encoded and exemplified predication---are governed by axioms and rules for which contemporary automated theorem provers lack tailored calculi. By contrast, reasoning in the shallow embedding of HOML in HOL remains closer to reasoning in the meta-logic, so that the resulting proof obligations can be discharged more easily using the rich tool stack already available for classical HOL (resolution and superposition provers, SMT solvers, model finders, Sledgehammer, etc.).

It is also worth noting that, as in PLM, foundational notions of mathe\-matics---including their modalised counterparts---can be identified in HOL as analysable objects that arise naturally, without requiring additional axiomatic postulates (with the exception of an axiom of infinity). This is hinted at in lines 32--35 of Fig.~\ref{fig:ChurchTTinHOML}, which follows Andrews' textbook \cite{Andrews02}, from which these definitions and abbreviations are drawn; we refer the reader there for further details. A thorough formalisation of this material within the \logikey\ framework remains, of course, a task for future work.

\section{Gödel's Theory Extended with Mathematical Notions}
\label{sec:godel}
Prior joint work with colleagues \cite{J52,J75} has formally established that the set of positive properties in Gödel's framework constitutes a modal set ultrafilter—and it has introduced the necessary modal machinery (modalised filter and ultrafilter definitions) to do so rigorously within the HOL meta-logic via the {\logikey} methodology. This even includes the careful distinction between modal ultrafilters defined on intensions versus extensions of positive properties \cite{J52}, and the distinction of actualist (varying domain) versus possibilist (constant domain) quantifiers in the involved postulates (see Fig.~\ref{fig:HOMLinHOL}).

Further mathematical notions will be required for the studies proposed here, covering foundational concepts such as natural numbers, equipollence, cardinality, and infinity. With these in place, one may investigate the formal effects of assuming such structures in interaction with Gödel's modal ontological theory, thereby reflecting his nuanced realist stance. The remainder of this paper offers an illustration of this direction, graphically captured in Fig.~\ref{fig:LogiKEy}; a full treatment is left for future work.

\subsection{Shallow embedding of HOML in HOL}\label{sec:embedding}

\begin{figure}[tp!]
\centering
\includegraphics[width=0.97\textwidth,frame]{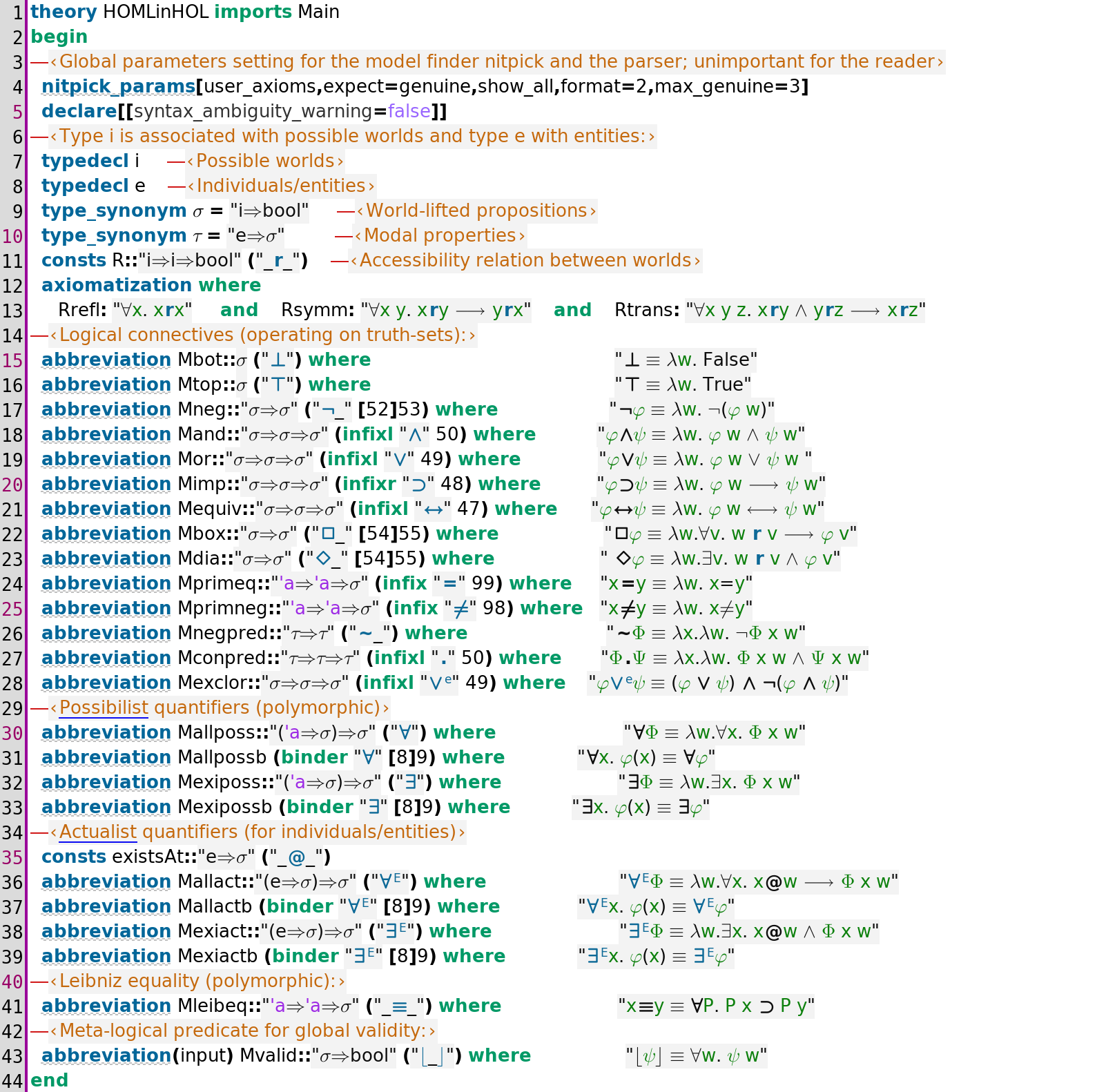}
\caption{Shallow embedding of higher-order modal logic (HOML) in the classical higher-order logic (HOL) of Isabelle/HOL utilizing the {\logikey} methodology. \emph{Reader's guide:} \(\iota\) is the type of (possibly non-actual) individuals, \(\mu\) the type of worlds, and \(o\) the HOL Booleans; the abbreviations \(\sigma:=\mu\to o\) and \(\tau:=\iota\to\sigma\) collect \emph{world-relativised propositions} and world-relativised (individual) properties, respectively (occurrences of \((\iota\to\sigma)\) in the figure may be read as \(\tau\)). HOML connectives are introduced as definitions on \(\sigma\): negation, implication, and the other propositional connectives are taken world-wise, and the box \(\Box\) is universal quantification over accessible worlds via the relation \(r:\mu\to\mu\to o\). \texttt{existsAt}\(:\iota\to\mu\to o\) (line 35) specifies which entities exist at which worlds and underwrites the actualist quantifiers $\mathbf{\forall}^E$ and $\mathbf{\exists}^E$, in contrast to the possibilist quantifiers $\mathbf{\forall}$ and $\mathbf{\exists}$ that range over the entire type. Two notions of equality appear: HOL identity \(=\) on the underlying carrier, and \emph{Leibniz equality} \(\equiv\), defined as the modalised statement that every property holding of one argument holds of the other. \emph{Global validity} \texttt{Mvalid}\(\,\varphi\) abbreviates \(\forall w.\,\varphi\,w\), the criterion under which HOML formulas are taken as theorems. Lines 12 and 13, postulating the axioms \texttt{Rrefl}, \texttt{Rsymm} and \texttt{Rtrans}—the only axioms introduced—configure the accessibility relation \(r\); making all three available, as here, specialises HOML to a higher-order analogue of modal logic \textbf{S5}, while subsets of these axioms recover modal logics \textbf{K}, \textbf{T}, \textbf{S4}, etc.}\label{fig:HOMLinHOL}
\end{figure}

\begin{figure}[tp]
\centering
\includegraphics[width=0.97\textwidth,frame]{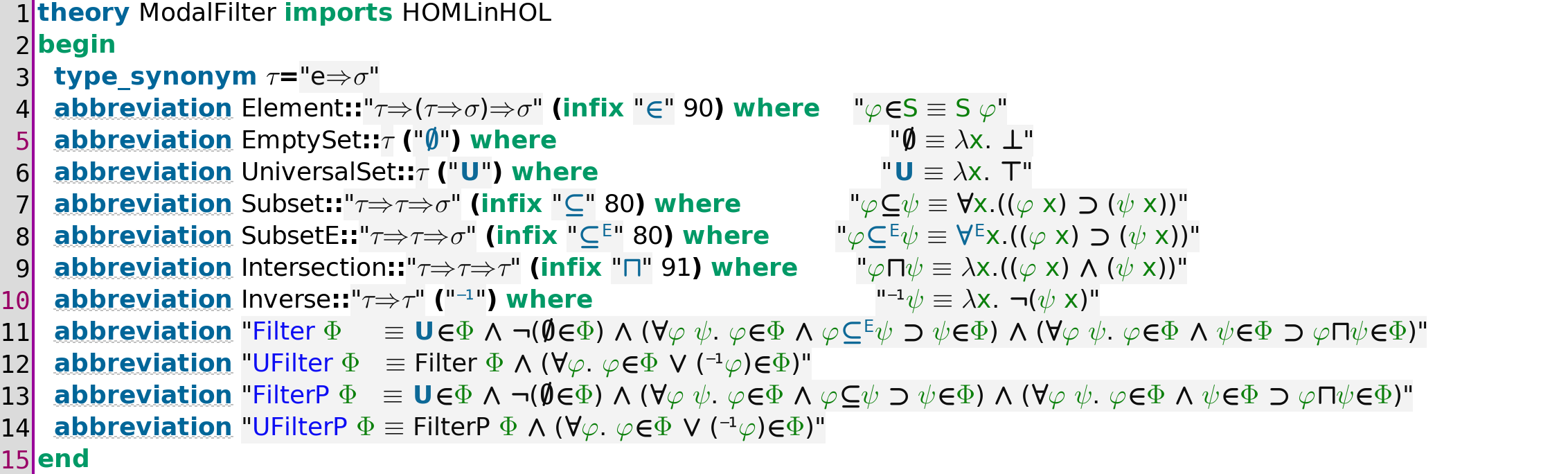}
\caption{Set filter and ultrafilter formalised for our modal logic setting. The types here can be read off as follows: a \emph{modal set} is a world-relativised predicate of type \(\tau\) (so it picks out, at each world, a subset of individuals), and a \emph{modal set of modal sets} accordingly has type \(\tau\to\sigma\). With this typing, \texttt{Subset} relates two modal sets (the inclusion is required to hold at every world), whereas \texttt{Element} relates a modal set to a modal set of modal sets---the difference in the two relations is forced by the difference in types and is exactly the modal analogue of the standard \(\subseteq\) versus \(\in\) distinction.}\label{fig:ModalFilter}
\end{figure}

The starting point for our illustration is the shallow embedding of HOML in
HOL, together with the definition of modal ultrafilters on top of it, in
exactly the form used in \cite{J75,J75afp}. These encodings (extracted from~\cite{J75}) are presented
in Fig.~\ref{fig:HOMLinHOL} and Fig.~\ref{fig:ModalFilter}, and they provide
the embedding of HOML in HOL that is the basis for what follows below; the
captions provide relevant information, and we refer the
reader to~\cite{J75} for full details.

\begin{figure}[tp]
\centering
\includegraphics[width=0.97\textwidth,frame]{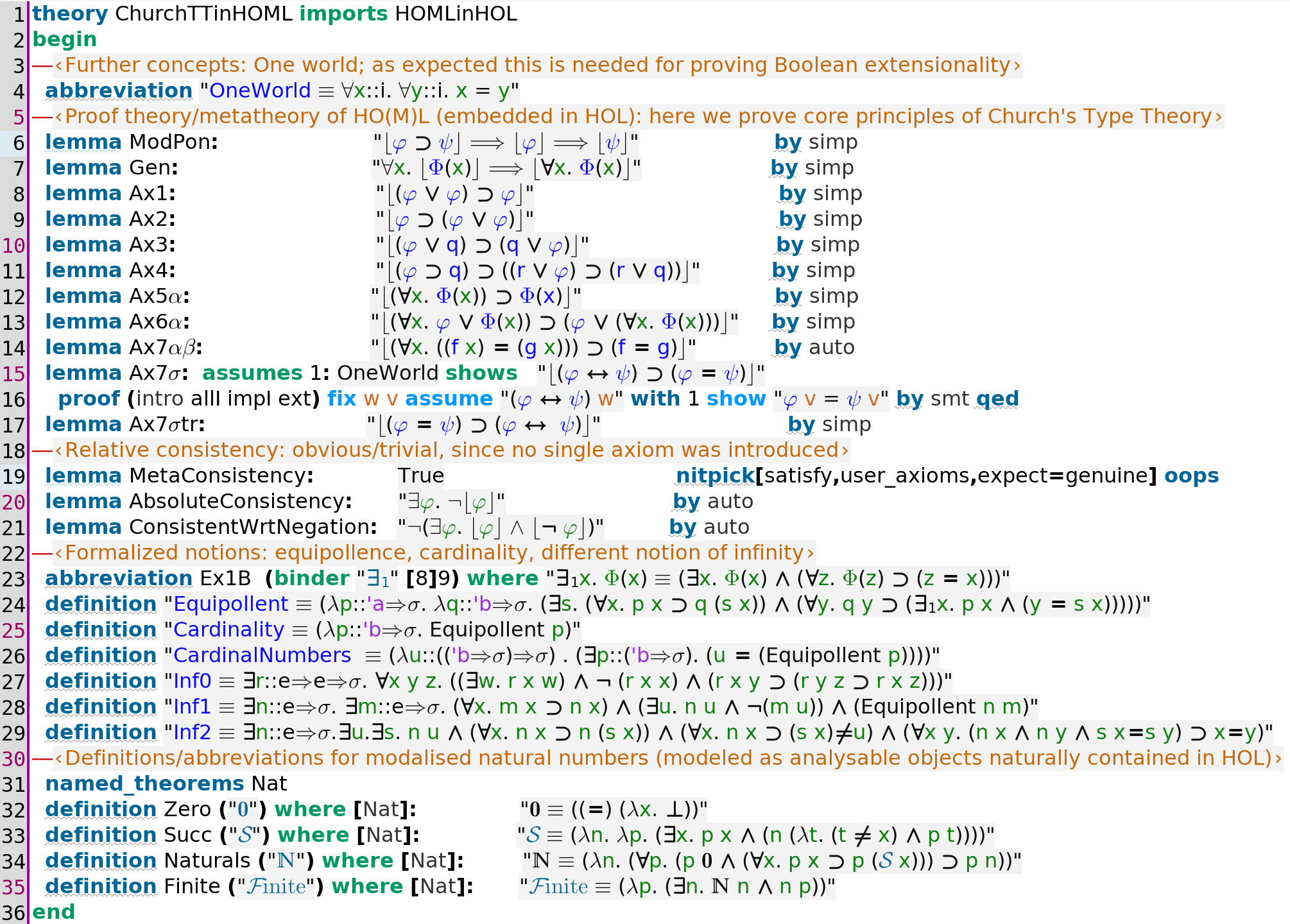}
\caption{Modalised versions of core (schematic) postulates of
Church's Type Theory are proven as lemmata, and further modalised
mathematical notions are provided as naturally embedded and analysable
concepts in HOL. Two notes may help to read the figure. 
First,
\texttt{CardinalNumbers}
(line~26) holds of a (modal) collection \(u\) when \(u\) is
the equipollence class of some witness \(p\). 
Second,
the successor operator \(\mathtt{S}\) on cardinals
(line~33) is \emph{not} a Church-numeral, fold-style
definition: given a cardinal \(u\), \(\mathtt{S}\,u\) is the cardinal
of \(p\cup\{*\}\) for some witness \(p\) of \(u\) and a fresh element
\(*\)---the standard Cantor--Bernstein successor in modal-set form.}
\label{fig:ChurchTTinHOML} 
\end{figure}

\subsection{Church's postulates at the HOML layer\label{sec:churchtt}}

Building on the embedding of HOML in HOL, Fig.~\ref{fig:ChurchTTinHOML} (lines 6--17) presents a study of the modalised versions of core postulates of Church's Type Theory \cite{Church40,B28},\footnote{Description and Choice are still left out here, but could easily be added.} an axiom system for HOL. The postulates of Church are lifted to the HOML layer and formulated as lemmata. All of these lemmata are shown to be derivable in HOML from the principles of the underlying meta-logic HOL—with one exception: the non-trivial direction of Boolean extensionality, i.e.~the lemma corresponding to Church's axiom~Ax7\(_{\sigma}\) (Fig.~\ref{fig:ChurchTTinHOML}, line~15). This is to be expected, since blocking Boolean extensionality is precisely the point of modal logic: unrestricted Boolean extensionality would permit arbitrary substitution of logically equivalent formulas even within the scope of modal operators, which is what modal logic is designed to prevent. Boolean extensionality can, however, be recovered in HOML under certain constraints. One option is to restrict the lemma to a particular fixed world—for instance, the actual world.\footnote{Since the embedding of HOML in HOL has access to the underlying Kripke structures at the HOL meta-layer, fixing and referring to particular worlds can be encoded straightforwardly (and has been done in prior work); the \logikey\ embedding technique thus scales naturally to hybrid logic.} A more drastic option, also illustrated here, is to establish Boolean extensionality for the degenerate case in which only a single world is assumed—a condition that would ultimately collapse the embedded HOML object-logic into full alignment with the meta-logic HOL.\footnote{One could reasonably object that the \texttt{OneWorld}
hypothesis ought not to be needed at all to recover the non-trivial direction
of Boolean extensionality at the HOML level. That it genuinely is needed can
be seen from a countermodel that \textsf{Nitpick} returns once the
\texttt{OneWorld} assumption is omitted. With a domain of two worlds
\(i_1, i_2\) and the total accessibility relation \(R\) (so that every world
sees every world), take \(\varphi\) to be false at both worlds and \(\psi\)
to be false at \(i_1\) but true at \(i_2\), and evaluate at \(w = i_1\). At
\(w = i_1\) the lifted equivalence \(\varphi \leftrightarrow \psi\) holds,
since both \(\varphi\) and \(\psi\) are false there; yet \(\varphi\) and
\(\psi\) differ at \(i_2\), so the HOL identity \(\varphi = \psi\)---which
requires agreement at \emph{every} world---fails. The lifted equivalence is,
by design, only a world-relativised statement asserting agreement at the
world of evaluation, and the local hypothesis it supplies at any single
\(w\) is therefore strictly weaker than the global HOL identity it is being
asked to support. Once the carrier of worlds is collapsed to a single world
by \texttt{OneWorld}, ``every world'' and ``the given \(w\)'' coincide, and
the lemma goes through.}

\begin{figure}[tp]
\centering
\includegraphics[width=0.97\textwidth,frame]{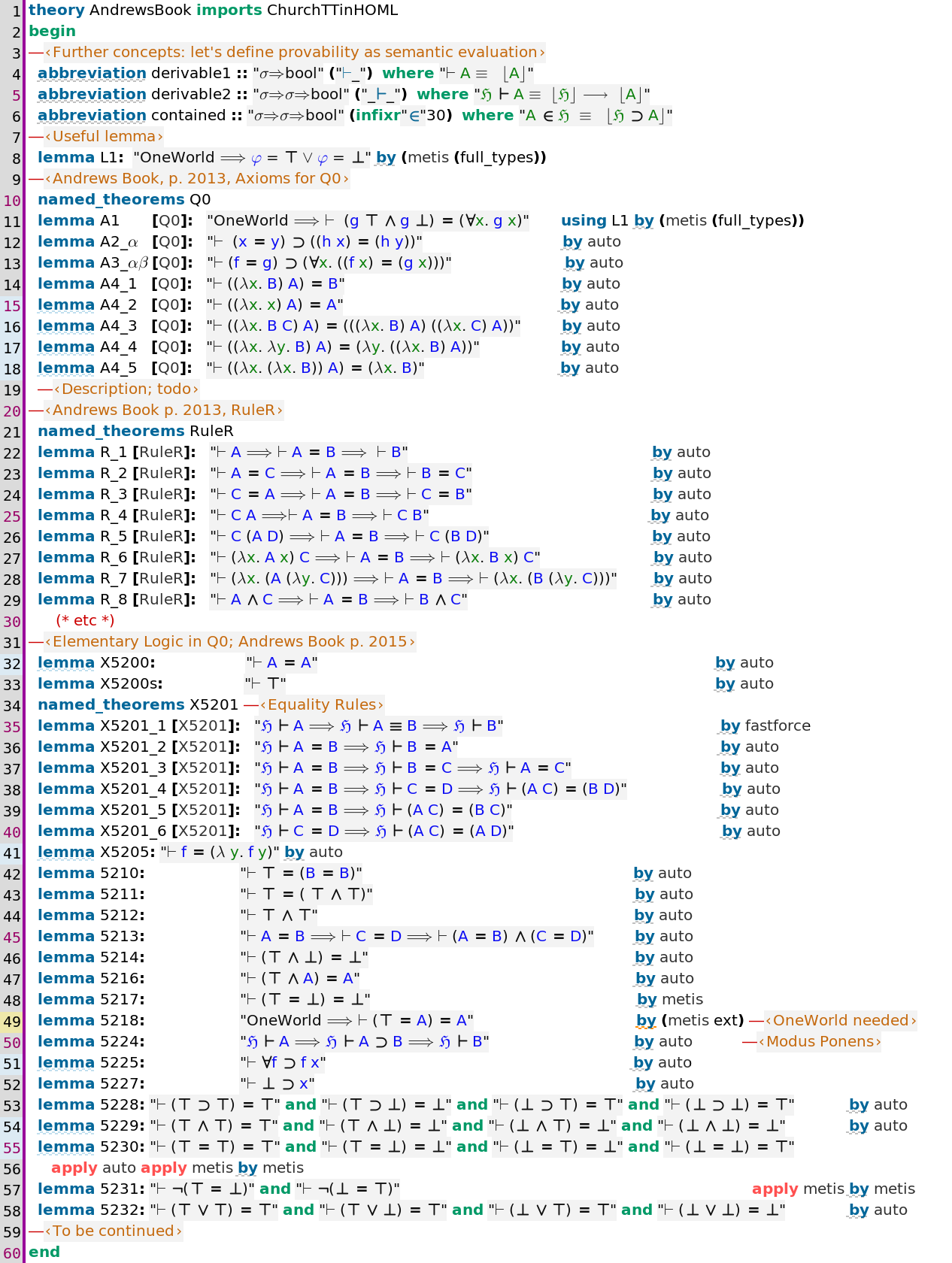}
\caption{Checking/verifying parts of Andrews' textbook~\cite{Andrews02} at the layer of HOML; the development is useful also for educational purposes. (This formalisation of Andrews' development is exploratory and partial: it is intended to illustrate the approach rather than to provide a complete treatment, which is left for an extended version of this paper.)} \label{fig:AndrewsBook}
\end{figure}

The above thus demonstrates that Church's prominent postulates for HOL are, with the expected exceptions, valid also for the embedded—and pragmatically more expressive—logic HOML. In principle, one could now proceed to encode and prove (largely automatically) the stepwise development of the foundations of higher-order logic and mathematics as carried out with great precision in Andrews' textbook \cite{Andrews02}; see Fig.~\ref{fig:AndrewsBook} below for an illustration.

\subsection{Modalised mathematical notions in HOML\label{sec:mathnotions}}

In Fig.~\ref{fig:ChurchTTinHOML}, however, we proceed differently. First, we use Isabelle's model finder \texttt{Nitpick} at the HOL meta-level to confirm the consistency of our development so far by generating a model; see line 19. Alternative formulations of consistency statements are shown in lines 20 and 21. All of these are of course entirely trivial, as the reader may readily observe: no axiom whatsoever—except for \texttt{Rrefl}, \texttt{Rsymm} and \texttt{Rtrans}—has been introduced in the embedding of HOML in HOL. Instead, the embedding shows, using definitions (resp.\ abbreviations) alone, that HOML can be identified as a substructure already naturally present within the meta-logic HOL.\footnote{Concretely, propositions of type \(\sigma:=\mu\to o\) are simply the world-indexed subsets of the (HOL) carrier of worlds. The set of positive properties, viewed as a filter, can in fact be given a topological reading: under suitable closure conditions (e.g.\ closure under arbitrary conjunctions and finite disjunctions) it forms a topology, and the modal ultrafilter perspective on Gödel's axioms (cf.~\cite{J52,J75} for the underlying filter/ultrafilter structure) then aligns with a familiar order-theoretic, indeed topological, picture---positive properties as ``large'' sets, modal necessity as a closure operator. This is a useful organising image rather than a load-bearing technical claim in the present paper, and the reader who prefers to read past it may safely take the embedding simply as world-indexed subsets of HOL.} This embedding of HO(M)L in HOL could of course be iterated further, giving rise to multiple reflection layers of embedded expressive logics within HOL—with the effect that each embedded object-logic remains open to formal study at its respective meta-level.

More relevant for the purposes of this paper, however, is the content presented from line 23 of Fig.~\ref{fig:ChurchTTinHOML} onward: an encoding of modalised variants of mathematical notions relevant to the work envisioned ahead. These include equipollence, cardinality, cardinal numbers, alternative notions of infinity, natural numbers, and finiteness; for details on these notions we refer the reader to Andrews' textbook \cite{Andrews02}. Importantly, all of these mathematical notions are introduced purely as definitions---or more precisely, as abbreviations for $\lambda$-terms in HOL---and thus require no additional axioms.

\subsection{Cardinality of positive properties\label{sec:cardinality}}

\begin{figure}[tp!]
\centering
\includegraphics[width=0.97\textwidth,frame]{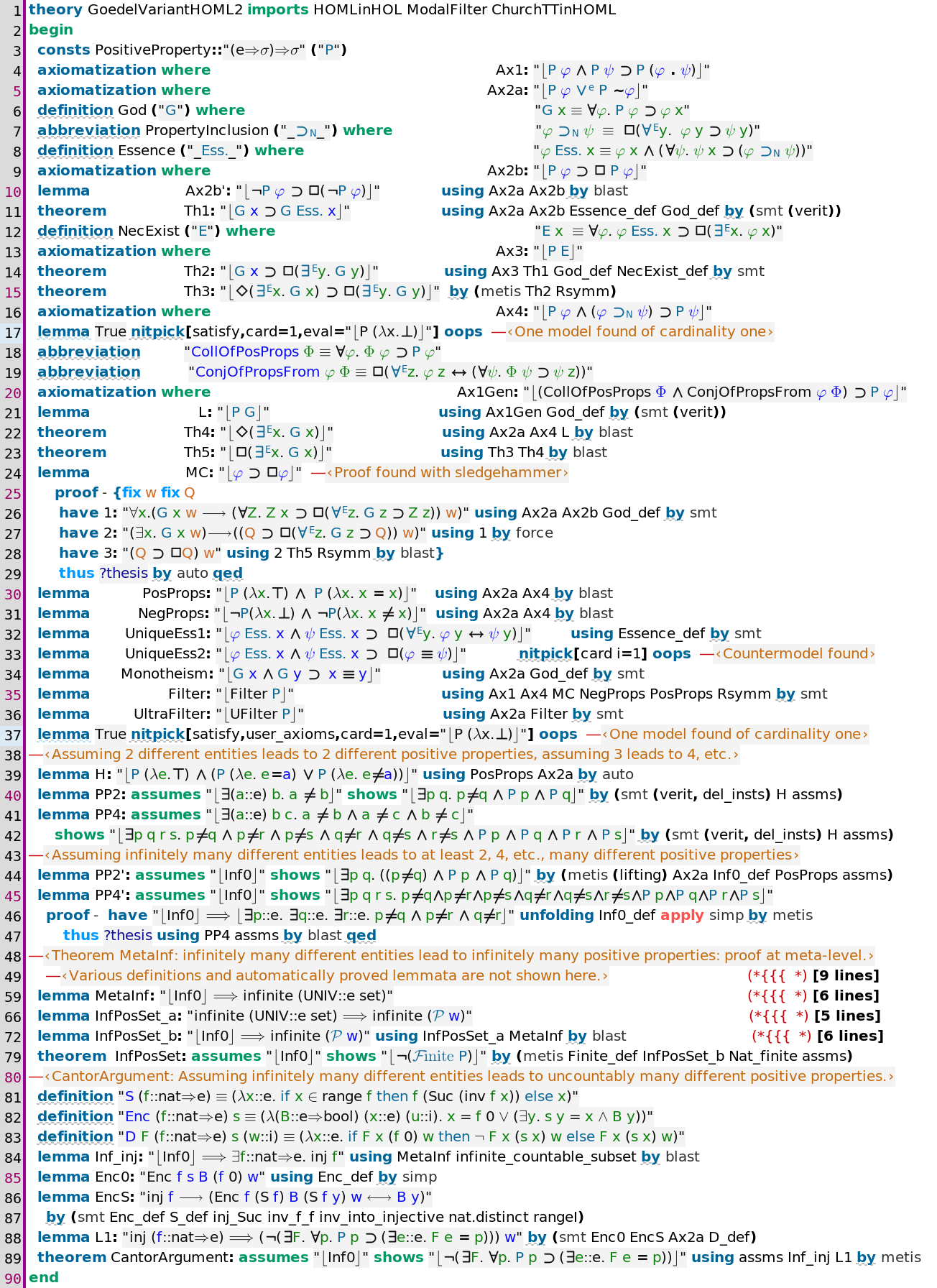}
\caption{The verification of Gödel's original modal ontological argument \cite{GoedelNotes} from \cite{J75}, enriched by a preliminary study on the cardinality of positive properties under the assumption of different existing (e.g., mathematical) entities.} \label{fig:GoedelVariantHOML2}
\end{figure}

Using the material introduced above, further experiments with Gödel's modal ontological argument are now possible, providing evidence for the effect mentioned earlier: namely, that assumptions about the existence of certain entities—mathematical ones, for instance—bear directly on the cardinality of the set of positive properties in Gödel's modal ontological theory.

This is illustrated in Fig.~\ref{fig:GoedelVariantHOML2}, starting from line~39. Lines up to 38 reproduce the experiments from Fig.~7 in \cite{J75} (with some interactive steps omitted for brevity), presenting a successful verification of the original version of Gödel's modal ontological argument as outlined in his 1970 manuscript \cite{GoedelNotes}. Lines 39--47 then show that assuming two existing entities within Gödel's theory yields two distinct positive properties; three existing entities yield four distinct positive properties; and so on.

\subsection{Uncountability via a modal Cantor argument\label{sec:uncountability}}

This line of reasoning is then extended to the infinite case. From line~48 onward, we verify that infinitely many distinct entities imply infinitely many distinct positive properties, as expected.\footnote{Our automated Isabelle proof of \texttt{InfPosSet}
(Fig.~\ref{fig:GoedelVariantHOML2}, line~79) establishes the existence of
infinitely many distinct positive properties by reasoning at the HOL
meta-layer about the HOML embedding---an injection from the natural numbers
into the entities is mapped into pairwise distinct positive properties, and the
resulting obligations are discharged via auxiliary lemmata (lines~59, 66, and~72) that
connect to the theory of natural numbers at the meta-layer---rather than arguing entirely
\emph{inside} the embedded HOML layer, using only our modalised mathematics
and the lifted Church postulates. There is nothing wrong with this proof in
principle---it arguably illustrates a virtue of \logikey---but constructing
an additional, pure HOML-level proof remains interesting future work.} 
Using a modalised and suitably constrained variant of the surjective Cantor theorem, we then strengthen this result and prove that the resulting infinite set of positive properties is in fact uncountable. The variant of Cantor's theorem used can be stated informally as follows: there is no surjective mapping \texttt{F} from the set \texttt{e} of entities into the set \texttt{P} of (modalised) positive properties (over entities). The corresponding theorem statement \texttt{CantorArgument} appears in Fig.~\ref{fig:GoedelVariantHOML2}, line~89, which depends on the crucial lemma \texttt{L1}. Note that the construction of the diagonal set, extracted here as an explicit definition $\texttt{D}$ which is used in the automated proof of lemma \texttt{L1}, is related to but significantly more involved in comparison to the standard proof (cf.~\cite[Fig.2]{J75} and~\cite{Lawvere2006}) of the surjective Cantor argument.

Concretely, we show—and formally verify in Isabelle/HOL—that assuming an infinite domain of entities, together with Gödel's axioms, gives rise to infinitely many—indeed uncountably many—distinct positive properties.

The next step in this project is to examine Gödel's generalised, third-order axiom Ax1Gen (Fig.~\ref{fig:GoedelVariantHOML2}, line~20) and its implications for the cardinality considerations of the present study---in particular the case where properties, viewed as mathematical objects, are themselves treated as existing entities. This line of inquiry will, however, inevitably encounter challenges arising from the hierarchy of simple types underlying HOL.

To state the objective of this work in more abstract terms: when the mathematical realist Gödel assumes that the objects of mathematics exist, then this rules out trivial finite and countable interpretations of positive properties—and with them, any finitely grounded or countable conception of God-likeness in his theory; cf.~also \cite{J76}.

In this ongoing work, the \logikey\ methodology again plays a central role, enabling systematic variation of the precise HOML under consideration—encompassing, for instance, variations between actualist and possibilist quantifiers, or between intensional and extensional interpretations of positive properties.

\section{Conclusion}
\label{sec:conclusion}

This position paper has reflected on the potential of combining logical pluralism with the \logikey\ shallow embedding methodology as a unifying framework for the computer-assisted study of foundational questions in mathematics, metaphysics, and theology. By embedding HOML in HOL, we have provided evidence that Church's postulates for HOL carry over, with one expected and well-understood exception, to the modal setting, and that Gödel's modal ontological argument, extended with modalised mathematical notions naturally contained in HOL as analysable objects, can be studied with considerable formal precision and a useful degree of automation. Our experiments further suggest that Gödel's mathematical realism may have direct and precise formal consequences: the assumption of mathematical entities forces the cardinality of positive properties beyond any finite and countable bound, with potentially far-reaching implications for the notion of God-likeness at the heart of his theory. The \logikey\ methodology has proven valuable throughout, enabling systematic variation of underlying logics and supporting (relative) consistency checking. A particularly promising methodological extension---hinted at in Sect.~\ref{sec:logikey}---is the integration of the deep-and-shallow embedding methodology recently developed for propositional and first-order modal logic~\cite{Benzmueller2025FaithfulDeepShallow,BenzmuellerKirchner2026FMLDeepShallow} into the HOML-based investigations pursued here: this would give the cardinality theorems on positive properties not only a mechanically verified semantic statement, but also a syntactic, deep-embedded counterpart; adequacy between the two would then be formally established as a theorem of HOL. Consolidating these findings, extending the analysis further, and maintaining careful attention to relative consistency constitute the central tasks ahead---with the broader aim of contributing to a rigorous, tool-supported dialogue between (meta-)logic, mathematics, metaphysics, and theology.

\paragraph{Acknowledgements.} We thank the reviewers for their
valuable comments, which significantly improved this paper. We are also
grateful to all \logikey\ collaborators over the past two decades, in
particular the co-authors of the referenced papers. Special thanks go to Ed
Zalta for fruitful discussions and to Dana Scott for his collaboration on
\cite{J75}. We also thank Claude.ai for dialogs that helped improve the paper
and shorten the novel proofs in Fig.~\ref{fig:GoedelVariantHOML2}. The work of
Luca Pasetto is supported by the Luxembourg National Research Fund (FNR)
(INTER/DFG/23/17415164/LODEX).

\printbibliography 

@article{J76,
  author       = {M{\"u}hlenbeck, Cordelia and Benzm{\"u}ller, Christoph},
  title        = {On the maximality of positive properties and modal
collapse in variants of Gödel’s ontological proof of God},
  journal      = {Logic and Logical Philosophy},
  url          = {https://apcz.umk.pl/LLP/article/view/55901},
  doi          = {10.12775/LLP.2026.007},
  year         = 2026,
  pages        = {1–21}
}

@Book{Andrews02,
  author =       "Peter B. Andrews",
  title =        "An Introduction to Mathematical Logic and Type
                  Theory: To Truth Through Proof",
  publisher =    "Kluwer Academic Publishers",
  year =         2002,
  edition =      "Second",
}

@InCollection{B28,
  author       =	{Benzm{\"u}ller, Christoph and Andrews, Peter},
  title        =	{{Church’s Type Theory}},
  booktitle    =	{The {Stanford} Encyclopedia of Philosophy},
  editor       =	{Edward N. Zalta and Uri Nodelman},
  url          =	{https://plato.stanford.edu/archives/spr2024/entries/type-theory-church/},
  year         =	{2024},
  edition      =	{{S}pring 2024},
  publisher    =	{Metaphysics Research Lab, Stanford University},
  OPTaddendum = {(substantive revision Jan 2024)},
}

@incollection{B9,
  Author =	 {Christoph Benzm{\"u}ller and Paulson, Lawrence C.},
  Title =	 {Exploring Properties of Normal Multimodal Logics in
                  Simple Type Theory with {LEO-II}},
  Year =	 2008,
  Booktitle =	 {Reasoning in Simple Type Theory --- Festschrift in
                  Honor of {Peter B. Andrews} on His 70th Birthday},
  OPTEditor =	 {Christoph Benzm{\"u}ller and Chad Brown and J{\"o}rg
                  Siekmann and Richard Statman},
  OPTIsbn =	 {978-1-904987-70-3},
  OPTKeywords =	 {own, Higher Order Logic, LEO Prover, Semantic
                  Embedding, Modal Logics},
  OPTNote =	 {(Superseded by 2013 paper in Logica Universalis)},
  Pages =	 {386-406},
  OPTPublisher =	 {College Publications},
  OPTSeries =	 {Studies in Logic, Mathematical Logic and
                  Foundations},
  OPTurl = {http://christoph-benzmueller.de/papers/B9.pdf},
  crossref = {E11},
}

@inproceedings{C26,
  Author =	 {Christoph Benzm{\"u}ller and Frank Theiss and
                  Paulson, Lawrence C. and Arnaud Fietzke},
  Booktitle =	 {Automated Reasoning, 4th International Joint
                  Conference, IJCAR 2008, Proceedings},
  Comment =	 {<a href="http://christoph-benzmueller.de/papers/2008_IJCAR_LEO-II.pdf">slides</a>},
  Doi =		 {10.1007/978-3-540-71070-7_14},
  Editor =	 {Alessandro Armando and Peter Baumgartner and Gilles
                  Dowek},
  Pages =	 {162-170},
  Publisher =	 {Springer},
  Series =	 {LNCS},
  Title =	 {{LEO-II} - A Cooperative Automatic Theorem Prover
                  for Higher-Order Logic (System Description)},
  OPTurl = {http://christoph-benzmueller.de/papers/C26.pdf},
  Volume =	 5195,
  Year =	 2008,
}

@inproceedings{C40,
  Author =	 {Christoph Benzm{\"u}ller and Woltzenlogel Paleo,
                  Bruno},
  OPTNote =	 {(Acceptance rate $\leq 28\%$)},
  Booktitle =	 {ECAI 2014},
  Comment =	 {<a href="http://christoph-benzmueller.de//papers/2014-ECAI.pdf">slides</a>},
  Doi =		 {10.3233/978-1-61499-419-0-93},
  Editor =	 {Torsten Schaub and Gerhard Friedrich and Barry
                  O'Sullivan},
  Pages =	 {93 -- 98},
  Publisher =	 {IOS Press},
  Series =	 {Frontiers in Artificial Intelligence and
                  Applications},
  Title =	 {Automating {G\"{o}del's} Ontological Proof of
                  {God}'s Existence with Higher-order Automated
                  Theorem Provers},
  OPTurl = {https://www.researchgate.net/publication/265050231},
  Volume =	 263,
  Year =	 2014,
}

@inproceedings{C44,
  Author =	 {Christoph Benzm{\"{u}}ller and Woltzenlogel Paleo,
                  Bruno},
  Booktitle =	 {Computer Science - Theory and Applications - 10th
                  International Computer Science Symposium in Russia,
                  {CSR} 2015,
                  Proceedings},
  Doi =		 {10.1007/978-3-319-20297-6_25},
  Editor =	 {Lev D. Beklemishev and Daniil V. Musatov},
  Pages =	 {398--411},
  Publisher =	 {Springer},
  Series =	 {LNCS},
  Title =	 {Interacting with Modal Logics in the {Coq} Proof
                  Assistant},
  OPTurl = {https://www.researchgate.net/publication/273201458},
  Volume =	 9139,
  Year =	 2015,
}

@inproceedings{C55,
  Author =	 {Christoph Benzm{\"u}ller and Woltzenlogel Paleo,
                  Bruno},
  OPTNote =	 {(Acceptance rate $\leq 25\%$)},
  Booktitle =	 {IJCAI 2016},
  Comment =	 {<a href="http://christoph-benzmueller.de/papers/2016-IJCAI-Poster.pdf">poster</a>, <a href="http://www.ijcai.org/Proceedings/16/Papers/137.pdf">proceedings-version</a>},
  publisher =	 {AAAI Press},
  Volume =	 {1-3},
  Editor =	 {Subbarao Kambhampati},
  Title =	 {The Inconsistency in {G{\"o}del's} Ontological
                  Argument: A Success Story for {AI} in Metaphysics},
  OPTurl =		 {http://www.ijcai.org/Proceedings/16/Papers/137.pdf},
  OPTOPTurl = {https://www.researchgate.net/publication/301295955},
  Isbn =	 {978-1-57735-770-4},
  Year =	 2016,
  Pages =	 {936-942},
}

@article{Church40,
  author    = {Alonzo Church},
  title     = {A Formulation of the Simple Theory of Types},
  journal   = {Journal of Symbolic Logic},
  volume    = {5},
  number    = {2},
  pages     = {56--68},
  year      = {1940},
  OPTurl       = {https://doi.org/10.2307/2266170},
  doi       = {10.2307/2266170},
  bibsource = {dblp computer science bibliography, https://dblp.org}
}

@book{Coq,
  author =	 {Bertot, Y. and Casteran, P.},
  publisher =	 {Springer},
  title =	 {{Interactive Theorem Proving and Program
                  Development}},
  year =	 2004,
}

@incollection{GoedelNotes,
  author =	 {K.~G\"odel},
  booktitle =	 {Logic and Theism},
  editor =	 {Sobel, J.H.},
  pages =	 {144-145},
  publisher =	 {Cambridge University Press},
  title =	 {Appendix {A}: Notes in {Kurt G\"odel's} Hand},
  year =	 1970,
  isbn =	 9781139449984,
}

@book{Isabelle,
  author       = {Tobias Nipkow and
                  Lawrence C. Paulson and
                  Markus Wenzel},
  title        = {{Isabelle/HOL} - {A} Proof Assistant for Higher-Order Logic},
  series       = {LNCS},
  volume       = {2283},
  publisher    = {Springer},
  year         = {2002},
  OPTurl          = {https://doi.org/10.1007/3-540-45949-9},
  doi          = {10.1007/3-540-45949-9},
  isbn         = {3-540-43376-7},
  bibsource    = {dblp computer science bibliography, https://dblp.org}
}

@article{J21,
  Author =	 {Christoph Benzm{\"u}ller and Paulson, Lawrence C.},
  Doi =		 {10.1093/jigpal/jzp080},
  Journal =	 {The Logic Journal of the IGPL},
  Number =	 6,
  Pages =	 {881-892},
  Title =	 {Multimodal and Intuitionistic Logics in Simple Type
                  Theory},
  OPTurl = {http://christoph-benzmueller.de/papers/J21.pdf},
  Volume =	 18,
  Year =	 2010,
}

@article{J23,
  Author =	 {Christoph Benzm{\"u}ller and Paulson, Lawrence C.},
  Doi =		 {10.1007/s11787-012-0052-y},
  Journal =	 {Logica Universalis (Special Issue on Multimodal
                  Logics)},
  Number =	 1,
  Pages =	 {7-20},
  Title =	 {Quantified Multimodal Logics in Simple Type Theory},
  OPTurl = {https://www.researchgate.net/publication/221677897},
  Volume =	 7,
  Year =	 2013,
}

@article{J30,
  Author =	 {Christoph Benzm{\"u}ller and Nik Sultana and
                  Paulson, Lawrence C. and Frank Theiss},
  Doi =		 {10.1007/s10817-015-9348-y},
  Journal =	 {Journal of Automated Reasoning},
  publisher =	 {Springer Netherlands},
  Number =	 4,
  Pages =	 {389-404},
  Title =	 {The Higher-Order Prover {LEO-II}},
  OPTurl = {https://www.researchgate.net/publication/280986731},
  Volume =	 55,
  Year =	 2015,
}

@article{J40,
  author =	 {Christoph Benzm{\"u}ller and Scott, Dana S.},
  title =	 {Automating Free Logic in {HOL}, with an Experimental
                  Application in Category Theory},
  volume =	 64,
  number =	 1,
  pages =	 {53--72},
  year =	 2020,
  OPTurl =       {http://doi.org/10.13140/RG.2.2.11432.83202},
  journal =	 {Journal of Automated Reasoning},
  publisher =	 {Springer Netherlands},
  doi =		 {10.1007/s10817-018-09507-7},
  OPTurl = {http://doi.org/10.13140/RG.2.2.11432.83202},
}

@article{J41,
  author =	 {Christoph Benzm{\"u}ller},
  title =	 {Universal (Meta-)Logical Reasoning: Recent
                  Successes},
  journal =	 {Science of Computer Programming},
  year =	 2019,
  volume =	 172,
  pages =	 {48-62},
  OPTurl = {http://doi.org/10.13140/RG.2.2.11039.61609/2},
  doi =		 {10.1016/j.scico.2018.10.008},
}

@article{J48,
  Author =	 {Benzm{\"u}ller, Christoph and Parent, Xavier and van
                  der Torre, Leendert},
  Journal =	 {Artificial Intelligence},
  Publisher =	 {Elsevier},
  Title =	 {Designing Normative Theories for Ethical and Legal
                  Reasoning: {LogiKEy} Framework, Methodology, and
                  Tool Support},
  Year =	 2020,
  Volume =	 287,
  Pages =	 103348,
  issn =	 {0004-3702},
  Doi =		 {10.1016/j.artint.2020.103348},
  OPTUrl =       {https://doi.org/10.1016/j.artint.2020.103348},
  OPTurl = {https://www.researchgate.net/publication/342146653},
  url_arxiv =	 {https://arxiv.org/abs/1903.10187},
}

@article{J51,
  Author =	 {Steen, Alexander and Benzm{\"u}ller, Christoph},
  title =	 {Extensional Higher-Order Paramodulation in {Leo-III}},
  journal =	 {Journal of Automated Reasoning},
  volume    = {65},
  number    = {6},
  pages     = {775--807},
  year      = {2021},
  OPTOPTurl       = {https://doi.org/10.1007/s10817-021-09588-x},
  doi       = {10.1007/s10817-021-09588-x}
}

@article{J52,
  Author =	 {Christoph Benzm{\"u}ller and David Fuenmayor},
  Journal =	 {Bulletin of the Section of Logic},
  Publisher =	 {Department of Logic, University of Lodz},
  Title =	 {Computer-supported Analysis of Positive Properties,
                  Ultrafilters and Modal Collapse in Variants of
                  {G\"odel's} Ontological Argument},
  OPTUrl =		 {https://czasopisma.uni.lodz.pl/bulletin/article/view/7976},
  Year =	 2020,
  Volume =	 49,
  Number =	 2,
  Doi =		 {10.18778/0138-0680.2020.08},
  OPTurl = {https://www.researchgate.net/publication/336742445},
  Pages =	 {127-148},
}

@article{J72,
  Author =	 {Alexander Steen and  Geoff Sutcliffe and Christoph Benzmüller},
  Title =	 {Solving Quantified Modal Logic Problems by Translation to Classical Logics},
  Journal      = {Journal of Logic and Computation},		  
  Doi =        {10.1093/logcom/exaf006},
  OPTEditor =	 {Andreas Herzig and Jieting Luo and Pere Pardo},
  Publisher =	 {Oxford University Press},
  volume       = {35},
  number       = {4},
  pages        = {1-23},
  year         = {2025},
}

@article{J75,
	title        = {Notes on {Gödel's} and {Scott's} Variants of the Ontological Argument},
	author       = {Christoph Benzmüller and Scott, Dana S.},
	year         = 2025,
	journal      = {Monatshefte für Mathematik},
	volume       = {208},
	pages        = {569–611},
	doi          = {10.1007/s00605-025-02078-x},
}

@article{J75afp,
  author       = {Christoph Benzmüller and Scott, Dana S.},
  title        = {Notes on Gödel's and Scott's Variants of the Ontological Argument ({Isabelle/HOL} dataset)},
  journal      = {Archive of Formal Proofs},
  OPTvolume       = {2025},
  year         = {2025},
  url          = {https://www.isa-afp.org/entries/Notes_On_Goedels_Ontological_Argument.html},
  }

@PhdThesis{KirchnerPhD,
  author = 	 {Daniel Kirchner},
  title = 	 {Computer-Verified Foundations of Metaphysics and an Ontology of Natural Numbers in Isabelle/HOL},
  school = 	 {Freie Universit{\"a}t Berlin},
  year = 	 {2022},
  doi = {10.17169/refubium-35141},
}

@InProceedings{Lambert60,
  author = {Karel Lambert},
  title = 	 {The Definition of E(xistence)! in Free Logic},
  year = 	 1960,
  booktitle = {Abstracts: The International Congress for Logic, Methodology and Philosophy of Science},
  publisher = {Stanford U. Press}}

@inproceedings{Lean2021,
  author       = {Leonardo de Moura and
                  Sebastian Ullrich},
  editor       = {Andr{\'{e}} Platzer and
                  Geoff Sutcliffe},
  title        = {The Lean 4 Theorem Prover and Programming Language},
  booktitle    = {Automated Deduction - {CADE} 28 - 28th International Conference on
                  Automated Deduction,  Proceedings},
  series       = {LNCS},
  pages        = {625--635},
  publisher    = {Springer},
  year         = {2021},
  OPTurl          = {https://doi.org/10.1007/978-3-030-79876-5\_37},
  doi          = {10.1007/978-3-030-79876-5_37},
  bibsource    = {dblp computer science bibliography, https://dblp.org}
}

@inproceedings{Obua2006HOLZF,
  author    = {Steven Obua},
  title     = {Partizan Games in {Isabelle/HOLZF}},
  booktitle = {Theoretical Aspects of Computing -- {ICTAC} 2006},
  series    = {LNCS},
  volume    = {4281},
  pages     = {272--286},
  publisher = {Springer},
  year      = {2006},
  doi       = {10.1007/11921240_19}
}

@unpublished{PLM,
	title        = {{Principia Logico-Metaphysica (Draft/Excerpt)}},
	author       = {Edward N. Zalta},
        OPTmonth        = 5,
	year         = 2026,
	note         = {Available at \url{https://mally.stanford.edu/principia.pdf} [accessed 24-May-2026]}
}

@misc{Paulson_ZFC_in_HOL_AFP,
  author    = {Lawrence C. Paulson},
  title     = {{ZFC in HOL}},
  howpublished = {Archive of Formal Proofs},
  year      = {2019},
  OPTnote      = {Formal proof development, \url{https://isa-afp.org/entries/ZFC_in_HOL.html}, accessed May 2026}
}

@techreport{R45,
  Address =	 {{DFKI Bremen GmbH, Safe and Secure Cognitive
                  Systems, Cartesium, Enrique Schmidt Str.\,5,
                  D--28359 Bremen, Germany}},
  Author =	 {Benzm{\"u}ller, Christoph and Paulson, Lawrence C.},
  OPTNote =	 {arXiv:0905.2435},
  Institution =	 {Saarland University},
  Publisher =	 {{SEKI Publications (ISSN 1437-4447)}},
  Series =	 {{SEKI Report SR-2009-02 (ISSN 1437-4447)}},
  Title =	 {Quantified Multimodal Logics in Simple Type Theory},
  OPTurl =		 {http://arxiv.org/abs/0905.2435},
  doi = {10.48550/arXiv.0905.2435},
  Year =	 2009,
}

@InCollection{Scott67,
  author = 	 {Dana Scott},
  title = 	 {Existence and description in formal logic},
  booktitle = 	 {Bertrand Russell: Philosopher of the Century},
  publisher = {George Allen \& Unwin, London},
  year = 	 1967,
  editor = 	 {R. Schoenman},
  pages = 	 {181-200},
  note = 	 {(See also: Philosophical Application of Free Logic, edited by K. Lambert. Oxford:OUP, 1991, pp. 28 - 48)}}

@incollection{ScottNotes,
  author =	 {D. Scott},
  booktitle =	 {Logic and Theism},
  editor =	 {Sobel, J.H.},
  pages =	 {145-146},
  publisher =	 {Cambridge University Press},
  title =	 {Appendix {B}: Notes in {Dana Scott's} Hand},
  year =	 1972
}

@book{Zalta1983,
  author    = {Edward N. Zalta},
  title     = {Abstract Objects: An Introduction to Axiomatic Metaphysics},
  series    = {Synthese Library},
  volume    = {160},
  publisher = {D. Reidel Publishing Company},
  address   = {Dordrecht, Boston, and Lancaster},
  year      = {1983},
  pages     = {xiii + 193}
}

@book{Zalta1988,
  author    = {Edward N. Zalta},
  title     = {Intensional Logic and the Metaphysics of Intentionality},
  series    = {Bradford Books},
  publisher = {The MIT Press},
  address   = {Cambridge, MA},
  year      = {1988},
  isbn      = {0262240270},
  pages     = {xiii + 256}
}

@unpublished{ZaltaUnifying2025,
	title        = {Unifying and Validating Some Ideas of
Kurt Gödel},
	author       = {Zalta, Edward N.},
    OPTmonth        = 5,
	year         = 2025,
	note         = {Awarded contribution to the 2025 Kurt Gödel Essay Prize; available online at \url{https://tinyurl.com/35t36nvz} [accessed 26-May-2026]}
}

@article{cs-LO-9301106,
  author       = {Lawrence C. Paulson},
  title        = {Isabelle: The Next 700 Theorem Provers},
  journal      = {CoRR},
  volume       = {cs.LO/9301106},
  year         = {1993},
  url          = {https://arxiv.org/abs/cs/9301106},
  bibsource    = {dblp computer science bibliography, https://dblp.org}
}

@inproceedings{mathlib2020, author = {{The mathlib Community}}, title = {The lean mathematical library}, year = {2020}, isbn = {9781450370974}, publisher = {Association for Computing Machinery}, address = {New York, NY, USA}, OPTurl = {https://doi.org/10.1145/3372885.3373824}, doi = {10.1145/3372885.3373824}, booktitle = {Proceedings of the 9th ACM SIGPLAN International Conference on Certified Programs and Proofs}, pages = {367–381}, numpages = {15}, location = {New Orleans, LA, USA}, series = {CPP 2020} }

@InCollection{sep-logical-pluralism,
	author       =	{Russell, Gillian and Blake-Turner, Christopher},
	title        =	{{Logical Pluralism}},
	booktitle    =	{The {Stanford} Encyclopedia of Philosophy},
	editor       =	{Edward N. Zalta and Uri Nodelman},
	howpublished =	{\url{https://plato.stanford.edu/archives/fall2023/entries/logical-pluralism/}},
	year         =	{2023},
	edition      =	{{F}all 2023},
	publisher    =	{Metaphysics Research Lab, Stanford University}
}

@book{E11,
  Isbn =	 {978-1-904987-70-3},
  Editor =	 {Christoph Benzm{\"u}ller and Chad Brown and J{\"o}rg
                  Siekmann and Richard Statman},
  Pages =	 {1--460},
  Publisher =	 {College Publications},
  Series =	 {Studies in Logic, Mathematical Logic and
                  Foundations},
  Title =	 {Reasoning in Simple Type Theory -- Festschrift in
                  Honor of {Peter B.~Andrews} on His 70th Birthday},
  Url =
                  {http://www.collegepublications.co.uk/logic/mlf/?00010},
  Year =	 2008,
}

@inproceedings{Benzmueller2025FaithfulDeepShallow,
  author    = {Christoph Benzm{\"u}ller},
  title     = {Faithful Logic Embeddings in {HOL} -- Deep and Shallow},
  editor    = {Clark Barrett and Uwe Waldmann},
  booktitle = {Automated Deduction -- {CADE-30} -- 30th International Conference on Automated Deduction, Proceedings},
  series    = {LNCS},
  volume    = {15943},
  pages     = {280--302},
  publisher = {Springer},
  year      = {2025},
  doi       = {10.1007/978-3-031-99984-0_16},
  note      = {Preprint: arXiv:2502.19311}
}

@inproceedings{BenzmuellerKirchner2026FMLDeepShallow,
  author    = {Christoph Benzm{\"u}ller and Daniel Kirchner},
  title     = {First-Order Modal Logic in {HOL}: Deep and Shallow Embeddings with Automated Faithfulness},
  OPTbooktitle = {{ARQNL} 2026: International Workshop on Automated Reasoning in Quantified Non-Classical Logics},
  year      = {2026},
  note      = {Submitted}
}

@article{Lawvere2006,
  author  = {Lawvere, F. William},
  title   = {Diagonal Arguments and Cartesian Closed Categories
             with Author Commentary},
  journal = {Reprints in Theory and Applications of Categories},
  number  = {15},
  pages   = {1--13},
  year    = {2006},
  note    = {Reprint, with commentary, of the 1969 original
             (Lecture Notes in Mathematics, vol.~92, Springer)},
  url     = {http://www.tac.mta.ca/tac/reprints/articles/15/tr15abs.html}
}

\end{document}